# Improving Device-Aware Web Services and their Mobile Clients through an Aspect-Oriented, Model-Driven Approach

GUADALUPE ORTIZ[1]

Quercus Software Engineering Group, University of Extremadura
Centro Universitario de Mérida
C/ Sta Teresa de Jornet, 38. 06800, Mérida (Spain)
Tf +34 924387068 Fax +34 924303782
gobellot@unex.es
and
ALFONSO GARCÍA DE PRADO[2]

Research Division ADT
Zaragoza (Spain)
Alfonso.garciadeprado@gmail.com

**Corresponding Author**: Guadalupe Ortiz, gobellot@unex.es,
guadalupe.ortiz@uca.es

**Abstract**
**Context**: Mobile devices have become an essential element in our daily lives, even for connecting to the Internet. Consequently, Web services have become extremely important when offering services through the Internet. However, current Web services are very inflexible as regards their invocation from different types of device, especially if we consider the need for them to be adaptable when being invoked from mobile devices.
**Objective**: In this paper, we provide an approach for the creation of flexible Web services which can be invoked transparently from different device types and which return subsequent responses, as well as providing the client's adaptation as a result of the particular device characteristics and end user preferences in a completely decoupled way.
**Methods**: Aspect-Oriented Programming and Model-Driven Development have been used to reduce both the impact of service and client code adaptation for multiple devices as well as to facilitate the developer's task.
**Results**: A model-driven methodology can be followed from system models to code, providing the Web service developer with the option of marking which services should be adapted to mobile devices in the UML models, and obtaining the decoupled adaptation code automatically from the models.
**Conclusions:** We can conclude that the approach presented in this paper provides us with the possibility of following the development of mobile-aware Web services in an integrated platform, benefiting from the use of aspect-oriented techniques not only for maintaining device-related code completely decoupled from the main functionality one, but also allowing a modularized non-intrusive adaptation of mobile clients to the specific device characteristics as well as to final user preferences.

**Author biographies:**

**Guadalupe Ortiz** completed her PhD in Computer Science at the University of Extremadura (Spain) in 2007. Since graduating in 2001 and for the following eight years, she worked as an Assistant Professor as well as a research engineer at the University of Extremadura's Computer Science Department. She has recently joined the University of Cádiz as Professor in the Computer Science Language and Systems department. She has published numerous peer-reviewed papers in international journals, workshops and conferences, and she has been a member of various program and organization committees of scientific workshops and conferences over the last years. Her research interests embrace aspect-oriented techniques as a way to improve Web service development in various fields, with an emphasis on model-driven extra-functional properties and quality of service, as well as their adaptation to mobile devices.

---

[1] Current address and affiliation: School of Engineering , University of Cádiz, C/ Chile, 1 11002 Cádiz (Spain). Tf. +34 956 015784. Fax. +34 956 015139. E-mail: Guadalupe.ortiz@uca.es

[2] Current affiliation: PhD Student at School of Engineering , University of Cádiz, C/ Chile, 1 11002 Cádiz (Spain).

**Alfonso García de Prado** graduated in Computer Science at the University of Extremadura in Spain and is currently pursuing a PhD at the University of Cádiz. For several years he has been developing programmer, analyst and consultant tasks for various international industry partners, his main focus being software development, evolution and management for large sport events. Over the last years he has paid special attention to research in the scope of mobile devices and Web services, in which sphere he has analyzed the advantages of using model-driven and aspect-oriented techniques, publishing several papers in the field.

## 1. INTRODUCTION

Mobile devices have acquired great prominence over the last years, with two and even three times more mobile users than Internet ones in some countries. The great amount of devices and their continuous use clearly illustrate the importance of access to mobile services [1].

From the service-side point of view, Web service developers have mainly focused on developing services which are designed to be accessible from desktop or laptop computers, creating a void in the sphere of their access from mobile clients – personal digital assistants (PDAs), mobile phones, etc., which are becoming very commonly used tools. In order to meet this requirement we have to bear in mind the type of device from which the service is going to be invoked. In this regard, developed clients will vary widely depending on the target device: there may not only be considerable differences between a mobile phone and a desktop computer, but also between clients developed for different types of mobile device. Furthermore, not only screen size, but also runtime capacity needs to be considered [2]. In this sense, we can say that desktop computers and laptops are the most powerful clients, allowing any type of computation, and are provided with screens which are expected to be large enough for information to be displayed clearly. PDAs and mobile phones, however, normally have less computational power and smaller screens than computers, as well as more reduced broadband access. Besides, they usually expose a reduced set of APIs (Application Programming Interfaces) due to the runtime environment they support.

On the other hand, from the client-side point of view, we have to take the device model into account: once we develop a client for a specific type of mobile device –i.e. a PDA-, we also have to account for large existing differences across the whole assortment of PDAs in the market. The information to be provided by the client will probably not vary much from one model to another, but the way it is displayed will, especially depending on screen size and shape. Besides, clients should also consider final user preferences: most mobile device screen settings can be personalized based on end-user preferences. Service clients should also take these into account, bearing in mind that the said preferences may vary over time and, consequently, so would resulting settings.

In this paper, we provide a solution for the creation of services which can be invoked from different types of device, particularly to adapt them to mobile clients, providing each one with the appropriate response. The solution is based on the transparent provision of information from the invoking device through the header of the sent SOAP message. The service will choose which piece of information to deliver depending on received information. In this approach *Aspect-Oriented Programming* (AOP) has been used to avoid scattered and tangled code [3, 4] across the system due to device adaptation, thus maintaining well structured and modularized code for the application. Furthermore, to save developers the need of learning an aspect-oriented language and to increase the implemented code's quality and reliability, device-related code has been automatically generated thanks to the use of a model-driven approach. In regards with the client-side, we also benefit from the use of aspect-oriented techniques in order to facilitate the adjustment of clients to diverse mobile devices' specific characteristics –such as screen size- as well as individual end user preferences.

The remainder of the paper is organized as follows: Section 2 gives background information on the technologies approached in this paper and provides a motivation of the problem, including a motivating scenario. Section 3 provides related work on the scope of the presented approach. Then, Section 4 outlines and compares the alternatives studied for creating Web services for multiple devices and describes implementation requirements on the selected alternative. Sections 5 and 6 describe the methodology to be used when developing service and client-side, respectively. Subsequently, section 7 illustrates the proposal following the procedure explained in the previous sections for the case-study presented in Section 2. Finally, Section 8 provides the approach evaluation and discussion, conclusions and our future line of work being mentioned in Section 9.

## 2. BACKGROUND AND MOTIVATION

In this section we provide a short background reference about the main technologies used in this paper, followed by a case-study motivation.

### 2.1. Background on Technologies

This paper provides a model-driven and aspect-oriented solution for Web service mobile clients, therefore this section is devoted to providing background information on the four technologies just mentioned. Since this paper provides a model-driven and aspect-oriented solution for Web service mobile clients, we will devote this section to providing background information on the four technologies just mentioned.

#### *2.1.1 Web Services*

Web services bring us one step further in the long way that object-oriented technologies and distributed platforms have come, having become an effective way to integrate third-party approaches on the Web, and collaborating in the client-server architecture replacement by peer-to-peer distributed architectures. They are often seen as applications accessible to other applications over the Web [5]. Such applications will be accessible via HTTP or SMTP, which facilitates message exchanges. Besides, services mainly use Simple Object Access Protocol (SOAP) as message format for communication. This message is composed of a mandatory body, where invocation parameters are included, and one or more optional headers.

#### *2.1.2 Model-Driven Development*

*Model-driven Development* (MDD) aims to promote the role of models, allowing us to focus on the system's conceptual design, delaying the decision of which implementation technology to use for a later step. Models may be used at multiple phases of development, from the initial system specification to testing. Each model will address one concern, regardless of the remaining issues involved in the system's development, thus allowing the separation of the final implementation technology from the business logic adopted by the system. Model-to-code transformations permit the system's automated development from models (see http://www.omg.org/mda/ for further information).

#### *2.1.3 Aspect-Oriented Programming*

*Aspect-Oriented Programming* (AOP) arises due to well-known problems detected in *Object-Oriented Programming* (OOP). One of the early objectives of OOP was the encapsulation and modularity of related data and methods which address common goals. However, experience has shown that it is impossible to model several concerns into a structured decomposition supported by current OOP practices.

To demonstrate this, let us consider a system for for geometric figures' representation and tracking. Suppose that there are two main concerns involved: the representation of shapes such as triangle, rectangle, and tracking of movements of a created figure (see [3] for further explanation). These are two transversal concerns which lack a logical decomposition of code structure by functionality, since any time we change the position of a figure vertex, we have to invoke the method to track its movement on the screen. Such concerns, which are often referred to as *crosscutting concerns*, result in code which is scattered and tangled all over the application. For example, code related to figure movement and update of their position on the screen is included within the same method, making it difficult to maintain the systems. AOP allows us to modularise crosscutting concerns by encapsulating them into meaningful independent units called *aspects*. In the previous example, we will have a method to change the figure's position and a separate aspect which will update it on the screen whenever the figure moves. Afterwards, a method to *weave* the aspect code with the original one is applied [6], method which interprets when every aspect code has to be performed in relation to the main functionality code.

Aspect-oriented techniques describe five types of element which can modularize crosscutting concerns: firstly, we have to define the *join point* model which indicates the points in the main system where new behaviours could be included. Then a way to indicate the specific join points for a particular system (i.e. Java Web services) needs to be defined; this is called the *pointcut*. Next, we ought to determine how we are specifying the behaviour to be included in the join points, that is, the functionality code which has to be performed whenever a join point is reached (the *advice*). We would then encapsulate the specified join points and their behaviours into independent units –*aspects*-. Finally, *weaving* of the new code with the original one has to be applied, such as pre-compiling and injecting aspects into the main system code [6].

### 2.1.4 Technologies for Mobile Devices

*Micro Java* (Java ME), formerly *Java Platform Micro Edition*, is the platform provided by Sun for running applications on mobile and other small or resource-constrained devices. Java ME can be used with two different configurations: *Connected Device Configuration* (CDC) and *Connected Limited Device Configuration* (CLDC). These specify the subset of Java APIs available for development. Besides, each configuration works with a different virtual machine: CLDC will work with the KVM virtual machine and CDC with CVM. On the one hand, Connected Limited Device Configuration is oriented to resource-constrained handsets and communicator-type devices; thus, it is designed for use with devices which are limited in their memory, graphical or computational capacity. On the other, Connected Device Configuration is optimized for resource-constrained devices, such as consumer products and embedded devices, therefore oriented to devices with more memory and computational capacity than CLDC.

### 2.2. Motivating Case-Study

In this section we are going to motivate the need for Web services which can provide several responses to the same request depending on the type of device from which invocations are performed and we will provide an example which will be used as the case-study for illustrating purposes. Nowadays, we can already download some applications for our mobile devices to access common websites (i.e. Google Mobile: http://www.google.com/mobile/), even Android-based devices have their own Google-based application. Another representative example would be an application to obtain information on stock market quotations. An economist speculating from his laptop would most probably be interested in as much information as possible: daily fluctuation of the company stock value, highest rate, lowest rate, a graph with last months' evolution, etc. However, when checking quotations from his phone, he most probably does not want to lose time and resources with graphs which are not always suitable for phones, but is interested in obtaining the current company stock quote as soon and up to date as possible. An online bookstore can also characterize the presented idea; we will provide this specific example in more detail, since it will later be used to illustrate our proposal.

Let us imagine we have an online bookstore which we have conventionally used to sell books through the Internet. So far, our customers have accessed the website through their PC or laptop, therefore with a screen wide enough to go through all available information on the book of their choice. For the purposes of the sale, the more information there is about the book, the more likely it is for the book to be purchased. Thus, the seller will try to show as much information as possible on required books on his website.

Let us now suppose that the business prospers, and the owner employs a staff member to visit prospective customers in order to increase sales. To enable the new employee to access information on the bookstore products wherever he is, the owner provides him with a high-end PDA, from which the employee will be able to access the bookstore website through the Internet and check any information on the books customers may be interested in. In this regard, it is not optimum to get as much information as we would get when invoking the service from a laptop or PC: for instance, the sales representative will not be interested in obtaining book reviews or example pages on the PDA, though he may be interested in having a look at the book index to check whether it covers the interests of the prospective customer. The same could be applied if our customer wants to access the bookstore through his mobile phone: he is definitely not interested in getting all available information on the book –he would most probably not be able to read the book index or its description comfortably-, but it will surely be appropriate to show basic information: title, author, publishing company and price.

In regards with the client-side, we can also consider that the device model, for instance a PDA, may have some particular characteristics such as screen size which may largely oscillate from one device model to another. The results of the invocation we selected earlier to be shown on a PDA can be displayed differently in accordance with different size screens in order to reach prospective buyers more easily, as market studies suggest. Besides, the end user may have selected specific setting for his device screen (such as background colour, type of font, etcetera). This may contribute positively towards the sale by displaying relevant information according to user tastes.

Thus, this common and straightforward example illustrates how we may require the same service from different types of device and how service invocation results should be adapted depending on the type of device being used. Besides, having to modify the mobile device client thoroughly depending on the device model is a redundant development and, furthermore, it would be impossible to create a client that meets end user preferences beforehand. In order to make these options available, we have to consider trying to avoid redundant code for the different device types and be aware that the implementation should be as transparent as possible in regards to the type of client.

As previously mentioned, this example will be used as the case-study for the remainder of the paper in order to illustrate our proposal's implementation and application. But prior to the description of our proposal, in the next section we are going to examine what other researchers have done in this area.

3. RELATED WORK

Regarding Web service adaptability, we mainly find work which deals with adaptability for service composition. For instance, in [7] and [8] we can see an example of provision of flexibility to Web services through the use of Aspect-Oriented programming. In this case, flexibility is interpreted as easier composition of Web services or the possibility of replacing some Web services by others, but these topics are out of the scope of this paper.

We can also find some proposals which deal with context adaptation. In this line of work, Menkhaus presents an architecture for decoupling user interfaces in Web applications from application logic [9]. Dockhorn et al. propose an architecture to support mobile context-aware applications through a publish-subscribe mechanism [10]. None of these approaches deals with the adaptation of Web service responses. Additionally, Pashtan et al. propose [11] the adaptation of Web applications' content depending on the device being used, but do not tackle Web service applications, which is the main focus in this approach. There is some more interesting work related to context, such as [12, 13, 14, 15 and 16]: for instance, Dorn et al. provide a hierarchy for context classification and description, as well as a query language and mechanism in order to provide and obtain contextual information, also through a publish-subscribe mechanism [14]. However, none of them provides a mechanism for using this context in the service-side to equate the answer to the client according to this context. On the other hand, Schomohl et al. provide context-aware mobile services [17], but they especially focus on the creation of location-based services, rather than the adaptation of service responses to client requirements.

The work from Keith et al. is more in accordance with our current approach. In [18] they present an approach for services to deal with client contextual information through a context framework. In their case, the context is always included in the client SOAP header as well as in service messages. This implies that not only services, but also clients have to process the context included in the header; however, the proposal does not explore how the client can deal with the received context. In our proposal the answer provided by the service is already adapted to client requirements, thus can be processed normally. Besides, their framework allows client context processing through the use of context plugins or context services. Context plugins have to be installed locally, which is improved by context services, available anywhere. A plugin and service have to be developed for each context and must be compatible with all services, which is extremely difficult and costly for developers. In our proposal, services do not need to be aware of context processing, which can be included at compile time through the use of aspects. Furthermore, since we provide a model-driven development, coding effort is diminished considerably. Finally, Song et al. extend the latest work in order to preserve the privacy of the client context in [19], yet do not provide any further advantage to our proposal.

Concerning the client side, the proposal from Zhang et al. provides the possibility of reengineering PC-based systems into a mobile product line by using a meta-programming technique that uses XVCL. In their approach, systems are firstly developed for PC environments and then evolved to mobile device platforms by generating specific components from generic metacomponents [20]. Alves' approach deals with existing variations in different mobile devices' models. He uses AOP to refactorize the variations and therefore decouple them from the core of the mobile application [21]. The idea of Blechsschmidt et al. consists of allowing the personalization of mobile device applications based on the end user profile [22]. For this purpose user information is collected and stored in XML-format files which are precompiled with the applications' core code in order for this information to be considered in the application during execution. Although all these proposals try to somehow deal with software characterization for mobile devices, none of them approaches the problem from a wide perspective where the development of the said software is considered from the service itself, and whereby device model and end user preferences are also contemplated.

Besides, the notion of maintaining personalization and graphical representation code decoupled from the core application is hardly tackled by these approaches.

Finally, a relevant piece of work is the one presented by D. Zhang in [23]. He provides an approach for Web content adaptation to meet the needs of users, fit characteristics of individual mobile devices, and adjust to dynamic contexts. His approach mainly focuses on Web applications and an interactive adaptation in which users have to take part during the invocation to obtain the information they want. There is no doubt this is valuable work which may be complemented with our approach, where we deal with software applications in general, not only Web-based ones, and adapt them at runtime to user preferences stored in the device as well as to the device model itself. Along the same line, we can find the work from Niederhausen et.al. [24], where a framework for Web applications adaptation is provided. The framework allows the developer to adapt Web application content depending on different adaptation concerns, such as device adaptation, through the use of what they call adaptation aspects. The base idea is quite similar to our approach, but again they focus the problem from the client-side, whereas we do it from the service side.

## 4. IMPLEMENTATION REQUIREMENTS OF DEVICE-AWARE WEB SERVICES

This section briefly describes and evaluates different alternatives for the creation of multiple device-aware Web services.

### 4.1 Alternatives for the Creation of Device-Aware Web Services

We assume that we have initially divided the different types of response expected from services into several groups, for example computers, CDC devices and CLCD devices, though this classification could be adapted to the evolution of coming technologies or systems' particular characteristics -it need not be CDC versus CLDC systems, it could be two CLDC devices with widely different characteristics, for instance. Once classified, our research into the possible approaches to getting device-aware services has led to four different alternatives:

- The first option for the creation of a service which personalizes the answer to received invocations depending on the type of device used is to provide clients with the possibility of adding a new parameter in the invocation. Subsequent information will be returned by the service depending on the parameter value.
- The second alternative is the addition of an optional tag in the invocation SOAP message header where the client can point out from which type of device he is performing the invocation, thus returned results will depend on the header value.
- Our third proposed option is to have different operations in the service for each type of device, but of course this would result in some redundant code in the developed service.
- Our last proposed solution is to create a set of façades for the different types of client [25]. The façades would invoke the existing Web service and adapt its result to the requesting device.

We have evaluated the four alternatives in order to develop the most efficient one. Table 1 shows a set of characteristics from the different proposals, which are described as follows:

1. Transparency: it will indicate whether the applied method requires the service developer to be aware of its adaptation for multiple devices or not when coding the service main functionality.

2. Consistency: this parameter indicates whether the service code still provides its functionality in a manner consistent to its original definition.
3. Non-Duplicity: with this parameter we will take into account whether or not we have to duplicate some service code.
4. Client-Awareness: with this parameter we can measure whether the client has to be aware of service multiple responses or if it could be unaware of it whilst still receiving an answer to his invocation.
5. Non-Intrusiveness: we will also consider whether it is necessary to include intrusive code in the service-side.

In the table, all these characteristics may take value 1 or 0: 1 means transparency, consistency, non-duplicity, client-oblivious or non-intrusiveness code are present and 0 shows that the said value is absent. We will now analyze the values shown in the table.

Table 1. Comparison among the different approaches

|  | Multi-parameter | SOAP Header | Multi-Operation | Façade |
|---|---|---|---|---|
| Transparency | 0 | 1 | 0 | 0 |
| Consistency | 1 | 0/1 | 1 | 1 |
| Non-duplicity | 1 | 1 | 0 | 0 |
| Client-Awareness | 0 | 1 | 0 | 0 |
| Non-intrusiveness | 0 | 1 | 0 | 1 |

1. The only option which is *fully transparent* for the service developer is the modification of the SOAP header. The remaining options imply that at least one parameter has to be present in the invocation or even that a new operation or service is needed.
2. The only proposal that could be regarded as not being *consistent* in appearance is the one where we modify the SOAP header, since the service is returning different information, apparently with the same entrance parameters, however it is indeed consistent to the message header provided. In the remaining approaches, consistency is also technically maintained, since the service returns different results for different invocations.
3. When we use either the SOAP header modification or the additional parameter approach we do *not* need to *duplicate* code in the service-side: the aspect will contain all code related to service adaptability. However, we cannot avoid some code duplication with the other two approaches.
4. The only option which is completely *client-aware* is the modification of the SOAP header, since the service may return a default value if nothing is included in the SOAP header to specify the invoking device. However, for the remaining options the client is at least aware of having to provide an additional parameter or invoke a different operation or service.
5. The proposals which do *not* insert any *intrusive* code in the service main functionality code are the SOAP header modification and façade ones, since the rest at least modify the signature of the operation to be invoked.

We have also implemented the motivating example making use of the four approaches and measured execution times. As shown in *Figure 1*, there are no relevant differences in execution times except for the façade implementation, thus –with the exception of façades– this should not influence our decision.

Figure 1. Average execution time per invocation to the four proposed alternatives

Based on this analysis we can infer that the SOAP header option is the most appropriate for the Web service environment, particularly due to its transparency, loose coupling and non-intrusiveness, which characterize these systems, as well as for its client obliviousness. A further analysis on each alternative can be found at [26, 27 and 28], where the initial phase of development of this research approach is shown. The following section describes the creation of device-aware Web services through the use of the SOAP header.

### 4.2 Implementation Details for the Creation of Device-Aware Web Services

We concluded that the best option for the selection of service responses is the addition of an optional tag in the invocation SOAP message header, where the client can point out from which type of device he is performing the invocation. We have found that this implies the following considerations for service implementation:

- In order to comply with a common WSDL service description, we will need the offered operation to return an element of a base class, which may be extended to different types of device or response. This means that if the service originally returned the full type class with the maximum possible information to be returned when invoked, now its definition will return the most basic class (in our case, where there are only two options -computer versus CLDC-, the full class would be the one returned to the computer –see *Info Class* in *Figure 2*- and the basic one is returned to the CLDC device –*Info_Base Class* in the illustration). Thus, we will have two complex type definitions –*base* and *extended*- the latter inheriting from the former, as shown in the named Figure.
- In order to discern from which device the received invocation has been pursued and, therefore, which information has to be provided in return, we will need to add a SOAP handler. This handler will intercept the service operation invocation and retrieve the tag value that indicates the type of device the invocation has been performed from. This tag will be available for service classes to read it.

Figure 2. Example of the stereotyping and splitting of a type definition class for mobile devices.

## 5. MODEL-DRIVEN DEVELOPMENT OF MOBILE AWARE WEB SERVICES

Once we know which the best alternative for the implementation of mobile-aware Web services is and what their implementation requirements are, we propose a model-driven aspect-oriented development approach. In this regard, in this section we will describe the model-driven methodology to be followed and the final aspect-oriented code obtained automatically in the driven transformations; specifically, subsection 8.1 will describe the *Platform-Independent Model* (PIM) and how it has to be marked for its coming transformation; subsection 8.2 will define the *Platform-Specific Model* (PSM) and how we will obtain it automatically from the platform-independent one and subsection 8.3 will provide the description of how the platform-specific model to code transformation will take place.

### 5.1 Platform-Independent Model

The system's platform-independent model will be represented as a class diagram. We assume that the services are represented as classes or interfaces, since this is the case in most Web service modelling approaches, such as [29] and [30]. Once we have the class diagram, we will have to mark it with its subsequent stereotypes to identify those relevant

elements for the device-aware PIM-to-PSM transformation, as we explain in the following lines:
- We will mark all classes to be Web services with <<*WebService*>>.
- Besides, we will also have to select those operations in the Web service which are to return different results depending on the invoking device. In this regard, <<*ws4md*>> has been applied to all methods which are going to return a different value for mobile or non mobile invoking devices.
- As a consequence of the previous decision, it will also be necessary to mark which complex type attributes returned by the <<*ws4md*>> operations are going to be returned to a mobile device. For this purpose we have also made use of <<*ws4md*>>, stereotype which has been applied to all classes defining a complex type. Besides, stereotype <<*cldc*>> has been applied to those attributes in the complex type definition class which have been selected to be returned to a mobile device.

Figure 3. Simple Mobile-Aware Web Service Platform-Independent Model.

In *Figure 3* we can see an example class (*Class1*) which has been stereotyped as a Web service; its offered method *Operation1* has been stereotyped with <<*ws4md*>>. *Operation1* will return different information depending on the invoking device; thus *Info* definition class has also been stereotyped with <<*ws4md*>>. In it, attributes *atributte1*, *atributte2* and *atributte5* have been stereotyped with <<*cldc*>>, which means that only that information will be returned for a mobile phone invoker; regular invoking devices would still receive full information.

### 5.2 Platform-Specific Model

The platform-specific model will mostly contain the same Web service class as the PIM, with the exception of its offered operation returning type, which will now be the base class. The main change in this model is that the complex type returned by the service offered method in the platform-independent model, has been converted into two classes: firstly, we will have a new definition class –*Info_Base*– containing those attributes returned to mobile invokers. Besides, *Info_Base* will be extended by *Info_Extended* with the remaining attributes, which will also include an additional method (*convertToBase*).

In order to automatically convert the platform-independent system model into the platform-specific one we have developed a plugin for Eclipse using *Model to Model Transformation Authoring* [31]. The plugin consists of the definition of a set of transformations from a UML model into another UML model:
- *Model2Model* will transform our source PIM into the target PSM.
- *Package2Package* will convert each package in the source model into a package in the target one.
- *Class2Class* will convert each class in the source model into a class in the service one.
- *Class2ClassBase* will create the new base class based on a <<*ws4md*>> type class, if required.
- *Class2ClassExtended* will create the new extended class based on a <<*ws4md*>> type class, should it be necessary.
- *Class2Operation* will create the *convertToBase* method in the extended class.
- *Operation2Operation* will convert each operation in the source model into another operation in the target one.

- *Parameter2Parameter* will convert each operation parameter in the source model into the corresponding target parameter.
- *Type2Type* will convert each type into the same one in the target model.
- *Property2Property* will convert each class attribute in the source model into the same one in the target model.
- *Property2PropertyExtended* will convert each class attribute in the source model into the same one in the target model, taking into account that it will now belong to the *Extended* class.
- *PrimitiveType2PrimitiveType* will convert each primitive type in the source model into the same one in the target model.

Once transformation rules have been defined, their Java code is generated automatically and the plugin is ready to be executed as an Eclipse application. *Figure 4* shows the result of transforming the platform-independent model in Section 4.1 into a platform-specific one.

Figure 4. Simple Mobile-Aware Web Service Platform-Specific Model.

### 5.3 Code

Finally, we assume that there are several platforms and tools which can generate code for Web services from their offered interface as well as for complex type definition classes, convertible into Java Beans (for instance several Eclipse plugins are used to that end). What is still to be generated is the necessary aspect code in order to adapt the service to the different invoking devices.

There are several aspect-oriented languages; particularly, AspectJ [32], one of the most well-known aspect-oriented languages for Java, has been selected for the aspect implementation in this approach due to its easy use, reliability and easy integration with development frameworks. AspectJ provides a large number of join points and pointcut designators; however, for our approach, only *method call* and *method execution* are relevant. This is because Web services are black box components and their visible parts are only their offered methods, therefore the visible join points are these method calls and executions. Specifically, since we are now injecting code in the service side, our join points will always be a *method execution* (*method call* would only be visible from the client side). Regarding the advices, the added behaviour can be executed before, after or around the pointcut, that is, we can add aspect functionality code before, after or around (before and after) method executions. In our approach, we make use of *around* pointcuts since we may need to replace the result returned by the method (depending on the invoking device) and therefore *wrap around* the method execution. The only aspect code which requires mastering AspecJ is the one which specifies the pointcut and defines the advice type; the functionality added in the advice is defined through the use of Java code, therefore we do not consider further details on AspectJ structure and syntax necessary. If the reader is interested, further details on AspectJ syntax can be found at [33] and an example of aspect code for the present approach is explained at [27].

To generate this code we propose developing transformations through the use of *Eclipse Project M2T (Model to Text)* [34] and specifically of *Jet Templates with Exemplars* [35], procedure which will be explained subsequently. With this transformation we will generate Aspect code which is compatible with any Java Web service implemented according to the designed system model. The same process could be followed for the SOAP handler; however, since each project will have the same code, its full code is provided and subject to reuse in any implemented system.

In order to create a model-to-code transformation using Jet with Exemplars we have to follow the following steps (further details can be found at [34]):

1. We have to provide the system with an exemplar file of the code type to be generated. In our approach, this would be an AspectJ aspect.
2. We also have to specify the tag structure of the prospective XML input file. In this case we have defined the UML models' XMI structure. In it we can see a reduced structure of a UML model, starting from the *package* element, followed by *packagedElements* and so on.
3. Then, we need to specify which parts of code in the exemplar file have to be replaced by specific tag values in the input file. For instance, in our proposal, operations intercepted in the pointcut will be replaced by operations offered by services which were stereotyped as device-aware ones. The template file is shown and explained in the coming paragraph after the fourth step.
4. Finally, given an input XMI model, we obtain the AspectJ file by simply executing the transformation.

The generated aspect code will be obtained once the transformation execution has retrieved all data required for the subsequent template, where generic names will be replaced by the specific service operation to be adapted to mobile devices and specific types to be returned. The set of lines (1) and (2) create two variables in order to store the input and output parameters of the intercepted Web service operation. Line (3) declares the package and line set (4) defines the particular aspect name making use of the service package name -<*c:get select="$xmi/package/@name" />*-, service class name -<*c:get select=$packagedElement/@name" />*- and adapted operation name -<*c:get select="$ownedOperation/@name" />*-. As we can see, we use JET tags and Xpath expressions for the retrieval of information from the XMI file representing the UML model, as well as for the inclusion of their values in the generated file (see [35] for further details on the use of JET Tags). Following the same XPATH terminology, we indicate in the pointcut –line set (5)– that we are going to intercept the operation to be adapted in the service, specifying its input parameters to be able to utilize them in the advice. The latter is provided in line set (6). Then we obtain the invoking device from the SOAP handler –line 7– and let the intercepted operation proceed with its execution in line set (8). Finally the result is converted to the invoking device type if necessary – line set (9) –, and the final result is returned – line (10).

```
(1) <c:setVariable select=
      "$ownedOperation/ownedParameter[not(@direction='return')]"
          var="ImputParams"/>
(2) <c:setVariable select=
     "$ownedOperation/ownedParameter[@direction='return']"
        var="OutputParams"/>

(3) package <c:get select="$xmi/package/@name" />;

(4) public aspect Aspect_<c:get select="$xmi/package/@name" />_<c:get
select=$packagedElement/@name" />_<c:get select="$ownedOperation/@name"/>{

(5) pointcut PC_<c:get select="$packagedElement/@name" />_<c:get select=
      "$ownedOperation/@name" /> (<c:iterate select=$ImputParams"
      var="myOwnedParameter" delimiter=", "><c:get select="substring-
      after($myOwnedParameter/type/@href, '#')"/> <c:get
      select="$myOwnedParameter/@name"/></c:iterate>):
            execution(* <c:get select="$packagedElement/@name"
            />.<c:get select="$ownedOperation/@name" />(<c:iterate
      select="$ImputParams" var="myOwnedParameter" delimiter=", "><c:get
      select="substring-after($myOwnedParameter/type/@href,
      '#')"/></c:iterate>))&&args(<c:iterate select="$ImputParams"
      var="myOwnedParameter" delimiter=", "><c:get
      select="$myOwnedParameter/@name"/></c:iterate>);
```

```
(6)   <c:get select="substring-after($OutputParams/type/@href, '#')"/>
      around(<c:iterate select="$ImputParams" var="myOwnedParameter"
      delimiter=", "><c:get select="substring-
      after($myOwnedParameter/type/@href, '#')"/> <c:get
      select="$myOwnedParameter/@name"/></c:iterate>): PC_<c:get
      select="$packagedElement/@name" />_<c:get
      select="$ownedOperation/@name" /> (<c:iterate select="$ImputParams"
      var="myOwnedParameter" delimiter=", "><c:get
      select="$myOwnedParameter/@name"/></c:iterate>){

(7)   deviceType device= MyHandlerClass.getDeviceType();

(8)   <c:get select="substring-after($OutputParams/type/@href,
      '#')"/>_Extended tmp = (<c:get select="substring-after
      ($OutputParams/type/@href, '#')"/>_Extended) proceed((<c:iterate
      select="$ImputParams"  var="myOwnedParameter" delimiter=", "><c:get
      select="substring-after($myOwnedParameter/type/@href, '#')"/>
      <c:get select="$myOwnedParameter/@name"/></c:iterate>);

(9)   if (device==deviceType.CLDC){
         <c:get select="substring-after($OutputParams/type/@href,
         '#')"/>_Base tmp2 = tmp.convertToBase();
         tmp=tmp2;}
(10)  return tmp;       } }
```

This template code is executed after the main template one's, in which, through a set of iterations, we move along the XMI file and specify that this code should be generated for all Web service operations stereotyped with <<*ws4mmd*>> in the UML model. Thus, as a result an aspect is generated for each operation which is also going to be offered for mobile devices.

## 6. LOOSELY COUPLED MOBILE CLIENTS ADAPTATION

Once chosen the SOAP header option for the creation of device-aware services, we have to deal with implementation consequences for the client-side. Besides, we also have to tackle how we are going to adapt clients to the specific device model and user preferences. In this section, we face these three facts: service invocation indicating device types in the SOAP header, dependences from the device model in the implementation and lastly final user preferences adaptation, in sections 6.1, 6.2 and 6.3, respectively.

### 6.1 Specifying the Type of Invoking Device in the Client-Side

Adding information to the SOAP header can be done simply by using a SOAP handler in a desktop/laptop client, which can utilize a full set of APIs and Java virtual machine. However, for mobile clients, it has to be taken into account that they neither have a full virtual machine nor a wide set of APIs –as explained in Section 2.1.4. For instance, Java ME with Connected Limited Device Configuration provides a reduced subset of Java APIs for development and works with the KVM virtual machine. Particularly, in the scope of Web services, this means that we have one API for service development -JSR 172 [36] - and that it is not possible to utilize handlers.

Bearing these considerations in mind, we made use of KSOAP2 libraries [37] in order to create the SOAP message for the invocation of services and to be able to include the header tag with the device specification in it. The code used to do so follows this explanation, where we can see that, once the SOAP envelope has been created, a new header with a new element –DEVICE- is created; DEVICE value is CLDC. Following this code, the SOAP body creation would appear, which is specific to the particular application.

```
(1) SoapSerializationEnvelope envelope = new
       SoapSerializationEnvelope(SoapEnvelope.VER11);
(2) Element[] headSoap = new Element[1];
(3) Element testEle = new Element();
(4) testEle.setName("IDENTIFICATION");
(5) Element testEle1 = new Element();
(6) testEle1.setName("DEVICE");
(7) testEle1.addChild(Element.TEXT, "CLDC");
(8) testEle.addChild(0, Element.ELEMENT, testEle1);
(9) headSoap[0] = testEle;
(10) envelope.headerOut = headSoap;
```

### 6.2. Tackling Client Dependences Resulting from the Device Model

Once we have developed a client specific to a type of device, there is an added constraint on the particular device model; especially for PDAs and mobile phones there is a large assortment of screen sizes, which implies that, if we want a better marketable product, results obtained in the invocation should be adapted to the most attractive layout according to screen size.

It is not efficient to create a new client for each different device model since it would imply a huge workload and it would most probably not be worth adapting clients to each specific device. Besides, it would be valuable to have the main functionality client code and the one originated by screen size restrictions completely separated rather than mixed.

With this purpose we propose the use of static aspect-oriented techniques. These will allow us to encapsulate all code related to those characteristics specific to the device model for the correct representation of results on the screen. In this regard, we will have, on the one hand, the general application code for a mobile device and, on the other, the aspects which will provide us with graphical representations of the data obtained in the invocation according to a specific device model. The aspect general code is shown below: as we can see, line (1) declares the aspect, and lines (2), (3) and (4) declare some required variables. Pointcut – the point to be intercepted in the main application, in this case the method *showInformation*, is declared in line (5). The new injected behaviour –the advice- goes from line (6) to (10), where we save the current display in the variable *screen* and create a new displayable element to be shown which is initialized through *myCanvasClass*. *myCanvasClass* is shown from lines (11) to (26): main values are initialized up to line (17), the method which displays the information on the mobile screen is shown from (18) to (23) -line (20) selects the type of font, line (21) sets it for the system and line (22) paints the ISBN number. Finally lines (24) and (25) allow the main screen to be shown again.

```
(1)public aspect ChangeDisplayMode {
(2) ISBN origen;
(3) myCanvasClass mycanvas_aspj;
(4) private Form screen;

(5) pointcut showInformation_p (ISBN ps, ItemBook_Base tmp):
   call(* showInformation(ItemBook_Base)) && target(ps) && args(tmp);

(6) void around(ISBN ps,ItemBook_Base tmp):showInformation_p(ps,tmp){
(7)     origin=ps;
(8)   screen=(Form) origin.display.getCurrent();
(9)     mycanvas_aspj=new myCanvasClass (tmp);
(10)    origin.display.setCurrent(mycanvas_aspj);     }

(11)    class myCanvasClass extends Canvas implements CommandListener {
(12)      private ItemBook_Base tmp;
```

```
(13)      myCanvasClass(ItemBook_Base _tmp) {
(14)          this.addCommand(new Command("Back", Command.BACK, 0));
(15)          this.setCommandListener(this);
(16)          this.setFullScreenMode(false);
(17)          tmp=_tmp;              }
(18)      protected void paint(Graphics g) {
(19)          [...]
(20)          Font T1=Font.getFont(Font.FACE_PROPORTIONAL,
                  Font.STYLE_BOLD, Font.SIZE_LARGE);
(21)          g.setFont(T1);
(22)          g.drawString(tmp.getISBN(), 40 , 10  ,
                  (Graphics.BASELINE | Graphics.LEFT));
(23)          [...]                  }
(24)      public void commandAction(Command c, Displayable d) {
(25)              origin.display.setCurrent(screen);       }
(26)  }        }
```

Once we have created the specific aspect for a particular device model, it can be compiled with the source application and deployed. Thus, we can predefine a set of aspects, specific to different device models, which can be reused for different client applications.

### 6.3. Tackling Client Dependences Resulting from End User Preferences

As previously introduced, there is an additional level of specialization in the client-side: personalization is based on the final user, who may define his tastes for his mobile device (font size, background colour, etc), once he is already using it.

In order to tackle this personalization, subsequent to the client installation in the device, we would propose the use of dynamic aspect-oriented techniques, which would permit the client personalization without having to recompile it and without the need of any intrusive code. However, since load time weaving [38] is not allowed in Java ME applications and we would have to modify the different loader in each specific device, we propose an alternative which allows the dynamic adaptation to user preferences, as explained below.

We have defined a set of characteristics which users can adapt to their preferences (font size and colour, for instance) and we have defined an xml file to describe them, such as:

```
<user_preferences>
        <font_colour v1= "..." v2= "..." v3= "..."/>
        <font_size value= "…"/>
</user_preferences >
```

The user can always change his preferences in the xml file, which is read by the aspect which shows the information on the device screen at runtime, thus adapting dynamically to end user tastes. The aspect will access settings provided by the final user through the use of JSR 75 [39]. As previously mentioned, this option provides us with the possibility of adapting clients to final user preferences as a way of obtaining better product sales by applying marketing techniques and strategies.

## 7 CASE-STUDY MODEL-DRIVEN DEVELOPMENT

This section will show how our case-study has been implemented using the presented approach in both service and client-sides, explained in detail in sections 7.1 and 7.2 respectively.

## 7.1 Case-Study Service-Side Development

Once the required elements for the PIM-PSM and PSM-Code transformations have been developed, they can be used in several projects, for instance in our case-study. In the following lines we explain the steps developers should follow to perform the model-driven development approach for the case-study system.

First of all, we have to model the class diagram for the system. In our system we have created the class *BookStore*; this class offers *getBookInfo* which may be invoked from different types of device and whose default return type is *BookInfo*.

Afterwards, based on the information provided for this system, the developer will add the necessary stereotypes to indicate which classes are going to be offered as Web services – *BookStore* is stereotyped with <<*WebService*>> in our case-study- and which of its operations are going to return different values depending on the invoking device type –*getBookInfo* is stereotyped with <<*ws4md*>>. Besides, <<*ws4md*>> is also applied to the complex type definition class returned by *getBookInfo* –*BookInfo*–. <<*cldc*>> stereotype has been applied to those attributes which are going to be returned when the invocation is performed from a mobile device. This model is shown in *Figure 5*.

Figure 5. Case-Study Mobile-Aware Web Service Platform-Independent Model.

Then, we have created a transformation configuration in Eclipse where we indicate that the source element is the model in the previous step and the target one is a new model project, and we select the transformation defined in our plugin to be applied.

The transformation execution will result in our platform-specific model being automatically generated from the platform-independent one. In this model, shown in *Figure 6*, we can see that the Bookstore service class definition model remains as it was in the platform-independent one, with only the operation return type changing into the new defined base class (*BookInfo_Base*). We can also see that we have a new class – *BookInfo_Base*–, containing those elements from *BookInfo* which are to be returned to a mobile device. Furthermore, this class is now extended by *BookInfo_Extended*, which only contains those complex type attributes which are not in the superclass and one extra method for the CLDC type conversion.

Figure 6. Case-Study Mobile-Aware Web Service Platform-Specific Model.

Next is code generation. As previously mentioned, we assume that both Web service and type class Java code can be generated from any tool. Then, to automatically generate AspectJ code to make our services device-aware, we will export the UML model in XMI format, which will be the input for our transformation. This step does not involve any difficulty since most UML tools are provided with an export-to-XMI option. The following step is to apply the defined transformation to the XMI file; this will result in the automatic generation of AspectJ code which may now be compiled in the service project. In the following lines we show the aspect code generated from the case-study platform-specific model in *Figure 6*. The aspect code will adapt service invocation results depending on the SOAP header content. In the following lines we can examine how exactly this is done through the aspect code, which is explained below:

```
(1) public aspect Adapting_getBookInfo{
(2)   pointcut p_getBookInfo(String ISBN) :
    execution(* BookStore.getBookInfo(String) &&args(ISBN);
(3)   BookInfo_Base around(String ISBN) : p_getInformation(ISBN){
```

```
(4)     deviceType device= MyHandlerClass.getDeviceType();
(5)     BookInfo_Extended tmp = (BookInfo_Extended) proceed(ISBN);
(6)     if (device==deviceType.CLDC){
(7)        BookInfo_Base tmp2 = tmp.convertToBase();
(8)        tmp=tmp2;}
(9)   return tmp;  } }
```

The aspect declaration is found in line (1). Line (2) specifies the pointcut, that is, the point in the service execution which we are going to intercept. As we can see, we are intercepting the execution of *getBookInfo* in *BookStore*. Lines (3) to (12) implement the advice: the code which we are executing in the point intercepted in the program's execution. As we can see in line (3), it is an *around* advice, which means that this code will be executed instead of the intercepted operation execution. The advice must return the same base class with which the intercepted service operation has been defined. Then, in line (4) we obtain the type of device which was indicated in the header of the received SOAP message invocation. Afterwards, the execution proceeds in line (5), that is, we allow the intercepted operation's execution to run as usual and we record the full resulting information in a standard result type. After this operation execution is concluded, depending on the value of the SOAP header device tag we convert the resulting object to previously defined type CLDC (lines 6 to 8), otherwise we let the full result be returned in line (9) and service execution continues normally. The definition of the different types will vary from one system to another depending on their specific settings.

### 7.2 Case-Study Client-Side Development

As previously explained, the type of device performing the service invocation has to be indicated in the SOAP header. The necessary code in order to do so and to make the invocation to our case-study is shown below: line (1) provides the location of the Web service for the SOAP message delivery, line (2) creates the SOAP message and then the code for the header creation –described in Section 6.1 – would come. Afterwards, lines (3) to (10) create the body of the SOAP message for serialization and delivery.

```
(1) HttpTransport httpt = new HttpTransport
        ("http://localhost:8080/AA_BiblioHan_WS/services/InfoBook");
(2) SoapSerializationEnvelope envelope = new
        SoapSerializationEnvelope(SoapEnvelope.VER11);
//[SOAP HEADER CODE]

(3) SoapObject request = new SoapObject("", "giveMeInformationOF");
(4) PropertyInfo pp = new PropertyInfo();
(5) pp.type = PropertyInfo.STRING_CLASS;
(6) request.addProperty("ISBN", tmpISBN);
(7) envelope.setOutputSoapObject(request);
(8) envelope.bodyOut = request;
(9) envelope.dotNet = true;
(10) envelope.encodingStyle = SoapSerializationEnvelope.ENC;
```

This SOAP message is sent to the service for invocation. The result of the invocation is shown in the right hand-side of *Figure 7*, where we can also see how different the result is when invoked from a laptop – left-hand side of Figure 7.

Figure 7. Bookstore invocation results from different types of device.

At a second level of specialization, we have to adapt the result representation depending on device models. Different mobile phones may require different letter sizes; for instance, *Figure 8* shows the adaptation of results for two completely different screen sizes. At the top of the illustration we can see a mobile phone with a large screen in which we have shown the service result with large characters on the left side device and small characters on the right one. The same thing has been done for a smaller screen device, whose results are shown in the lower part of the illustration. We can clearly appreciate how large characters are more suitable for large screen devices and small characters for smaller ones. Let us remind the reader that this has been done through the use of an aspect so that the main functionality code remains untouched, regardless of the invoking device features.

Figure 8. Bookstore invocation results from different device models using different sizes and from the same device using final user preferences for background and font colour.

Finally, the result has also been adapted to user preferences, also shown in *Figure 8*: the same result from the service invocation is shown with the standard set of colours in the device, using the same background colour as the device body and the Spanish flag colours. This shows how easily results can be adapted to predefined user preferences.

## 8. EVALUATION AND DISCUSSION

In this section we are going to evaluate and discuss the described approach. Several aspects are subject to evaluation:

Firstly, it could be thought that execution times were affected by the use of aspect-oriented techniques. Nevertheless, AOP weavers, specifically AspectJ ones, have evolved considerably and the latter community aims for the performance of AspectJ implementation to be on par with the same functionality coded in Java. In spite of this assertion, we have measured execution times of the motivating example, implementing it both using the presented approach with AspectJ versus only using Java. Measurements have been taken by performing 100, 1000, 50000 and 500000 invocations to the services a minimum of five times. Invocation results reveal that execution times tend to be rather similar, with slight differences of hundreds of milliseconds (further comparative performance analysis and Figures representing execution times can be found at [28]). Thus, with the use of aspect-oriented techniques we improve our systems with well modularized code for the system and no intrusiveness at all for the original Web service code. Besides, services can be easily extended to return different information depending on customer requirements, thus improving system code maintenance and evolution.

Regarding the use of the SOAP header to indicate the client side invoking device, and service side use of handlers to check which information is included in the named header, it may be thought that it can also damage invocation performance. Even though the use of the SOAP header to provide information related to service management is common practice and it is assumed that the developer has to make a decision in the compromise between modularity and performance, we have made some execution times measurements to assure ourselves that performance is not vitally damaged. In fact, performance measurements implementing the approach with and without SOAP –adding an extra parameter in the invocation - also both with Java and AspectJ have been added to the previous ones, measurements have been taken by performing 100, 1000, 50000 and 500000 invocations to the services a minimum of five times. Figure 9 represents the results of average times of each of these sets of executions for the aspect and non aspect-oriented implementations in a logarithmic scale. In order to make the results clearer we have also represented, in Figure 10, the average response time of one invocation for each of the described sets of invocations in both implementations. As we can observe in the illustration, execution times tend to be rather similar, with slight differences of hundreds of milliseconds; therefore neither the use of AOP nor the use of the SOAP header and handler imply great differences in our application's performance.

Figure 9. Execution times for sets of invocations to Java (J) and AspectJ (AJ) implementations, both introducing the invoking device in the SOAP header (Header) or through a new parameter (Param).

Figure 10. Average execution time per set of invocations to Java (J) and AspectJ (AJ) implementations, both introducing the invoking device in the SOAP header (Header) or through a new parameter (Param).

In regards with the model-driven development, the learning curve due to the incorporation of AOP code is avoided for the service developer, as well as saving the added workload of making services device-aware. At the same time, we prevent possible coding errors and facilitate quick maintenance of aspect-oriented classes by providing new transformation files when required. Concerning the client side, the use of aspects for the separation of the client application's main functionality from its representation depending of the device model helps us maintain our system code well modularized, allowing us to reuse the application in different models simply by changing the representation aspect, whose skeleton code can also be generated automatically, if necessary.

On the other hand, the developer will have to define the different types of object which will extend base classes and decide which one will be returned to each type of invoking device. This is obviously necessary if we want to return different results depending on the device and does not imply excessive workload when coding the system. For further personalization in the client side, the use of aspects and an xml file allows us to easily add or remove these features based on the final user, regardless of the application functionality.

Finally, it is our desire to mention that, even though screenshots shown in this paper are captured from mobile emulators, the case-study Web service has been tested using two real mobile devices -a Samsung Omnia HD and a Samsung Galaxy handset. These two devices were chosen due to the fact that they have different operative systems (Symbian and Android, respectively) so as to prove that the presented approach is compatible with diverse devices.

## 9. CONCLUSIONS AND FUTURE WORK

The approach presented in this paper provides us with the possibility of following a model-driven development of mobile-aware Web services in an integrated platform. Thanks to this fact we rise to one of the current challenges of software development, which is the alignment of current output with subsequent steps input in a development process. In this approach everything has been integrated with Eclipse or RSA plugins in the same platform, so that the full development process can be easily followed from initial platform-independent model to final implementation code.

Besides, as we previously introduced, the approach is perfectly extensible to different types of result the service has to return (not only mobile phone versus computer). Both platform-independent and platform-specific models are easily extensible for this purpose with the only addition of a stereotype for each new type of result to be returned. The new aspect required for the modularized and non-intrusive addition of code will be automatically generated by previously described model-to-model and model-to-text transformations.

Furthermore, the use of aspects only benefits the approach, allowing us to maintain device-related code completely decoupled from the main functionality one. In this regard, a last minute change of requirements would only imply an aspect addition or deletion, without the need to regenerate the full system. Aspects' performance has been measured and compared with an equivalent implementation without aspects, showing that aspect inclusion does not impact negatively on system performance. What is more in regards with client-side applications, aspects allow a modularized non-intrusive adaptation of the latter to the specific device characteristics in which they are going to be deployed as well as a dynamic adaptation to final user preferences.

On the other hand, the present approach deals with the development of systems from a platform-independent model, assuming that the knowledge of the computational-independent model is known and applied by the developer when marking the PIM. A proper definition of the Computational-Independent Model and its automated processing for the inclusion of stereotypes in the PIM model would no doubt enrich the presented approach, and it is planned for our future research.

Finally, another extension which would improve the presented work would be to include the possibility for final device users to choose the type of response they receive as part of the dynamic preferences which can be changed at runtime. To date, the application developer knew intuitively which data were relevant for display on mobile devices. However, this is sometimes subjective and depends on user requirements. In this regard, we plan for our future work to let the final user decide if he wants to receive full or mobile-adapted information.

## ACKNOWLEDGMENTS

This work has been developed thanks to the support of MEC under contract TIN2008-02985 and MEC research grant *José Castillejo*.